\documentclass[aps,amsmath,amssymb,nofootinbib]{revtex4} 
\usepackage{graphicx} 
\usepackage{amsmath}
\usepackage{amsfonts,amsbsy}
\usepackage{amssymb}

\def\be{\begin{equation}}
\def\ee{\end{equation}}
\def\bea{\begin{eqnarray}}
\def\eea{\end{eqnarray}}
\newcommand{\ud}{\, \mathrm{d}}
\newcommand{\xt}{{\mathbf{x}}}
\newcommand{\yt}{{\mathbf{y}}}
\newcommand{\rt}{{\mathbf{r}}}
\newcommand{\bt}{{\mathbf{b}}}

\newcommand{\kt}{{\mathbf{k}}}
\newcommand{\nabt}{\boldsymbol{\nabla}}
\newcommand{\tr}{\, \mathrm{tr} \, }

\begin{document}

\title{\bf 
High order cumulants of the azimuthal anisotropy in the dilute-dense limit: Connected graphs.  
}

\author{Vladimir Skokov}
\email{Vladimir.Skokov@wmich.edu}
\affiliation{Department of Physics, Western Michigan University, Kalamazoo, MI 49008, USA}

\begin{abstract}
We analytically compute higher order cumulants  of the azimuthal
anisotropy, $c_2\{2m\}$, and corresponding $v_2\{2m\}$ at high transverse momentum
in the dilute-dense limit. The dense target is considered in the framework
of 
the McLerran-Venugopolan model.   
The absolute values of the harmonics  $v_2\{2m\}$ of the azimuthal anisotropy 
are approximately equal, 
$|v_2\{2m\}| \approx |v_2\{2m'\}| $,  for large $m$ and $m'$.
However, the harmonics with order $2m=4n$ are complex. 
We argue that this is a generic property of connected graphs,
which remains true in the dense-dense limit.  
\end{abstract}

\maketitle

\section{Introduction}

Large azimuthal asymmetries  observed in p+Pb  collisions at
the LHC~\cite{pPb_ALICE,pPb_ATLAS,pPb_CMS} and in d+Au collisions at
RHIC~\cite{dAu_RHIC}
have usually been described by hydrodynamics~\cite{hydro} or the ``glasma graph''~\cite{pAridge} computed in the framework 
of the Color Glass Condensate (CGC). Both approaches resulted in a  fairly good description of the 
data for two particle correlations. 
The origin of the asymmetries is, however, very different: in hydrodynamics the asymmetry is related
to the azimuthal anisotropy for a single particle, 
while in the Color Glass Condensate~\footnote{We 
note that the ``glasma graph'' does not 
exhaust all the sources of the azimuthal anisotropy within the CGC framework, see e.g. Refs.~\cite{KL,BFKL,AniCGC,AniMV,AniC4}.} 
it is related to the correlation 
of gluons in the dense target 
(and the dense projectile in case of the dense-dense limit).

Higher order cumulants of azimuthal anisotropy and associated harmonics  are 
a very sensitive probe of {\it collectivity} in the system because they enhance non-trivial 
$m$-particle correlations. These correlations naturally appear in a hydrodynamical description of 
high-energy p-A and A-A collisions. However, we warn the reader from equating hydrodynamics and collectivity, 
because intrinsic correlations between $m$-particle
and  apparent
collectivity may arise owing to 
many-particle dynamics of partons in the target wave function. 
As an example, we note that 
the gluon saturation is already a genuinely 
collective/many-particle phenomena.   

Nonetheless, we show herein that 
higher order cumulants are capable of narrowing down the list of 
models used to describe high-energy  hadron collisions. 

In this article we consider the 
dilute-dense limit and compute higher order 
cumulants of the azimuthal anisotropy.  
We show that fully connected graphs, 
which are topologically equivalent to the glasma graph
of the Color Glass Condensate, 
produce positive four-particle cumulants, i.e.  {\it complex} $v_2\{4\}$. 
We also show that the {\it absolute} value of the harmonics $v_2\{m\}$ for large $m$ 
is non-zero and approximately independent of $m$.

\section{Calculations of high order cumulants of azimuthal anisotropy}
\subsection{S-matrix}
In the dilute-dense limit, the projectile is modeled as a collection
of partons scattering off the classical field of the target. We treat
the target in the semi-classical approximation following the McLerran-Venugopalan model.
Scattering of a parton  off the target is quantified by the S-matrix
\begin{equation}
{\cal S}_1(\rt,\bt) \equiv \frac{1}{N_c}\tr V^\dagger(\xt)\,
V(\yt)~, \label{Eq:S_1}
\end{equation}
where the dipole radius and the impact parameter are $\rt \equiv \xt-\yt$ and $\bt \equiv \frac12(\xt+\yt)$ respectively.
$V(\xt)$ is the Wilson line describing propagation of the parton in the field of the target
\be \label{eq:V_rho}
V(\xt) = \mathbb{P} \exp\left\{ ig \int \ud x^-  
A^+(x^-,\xt) \right\}. 
\ee
We restrict our consideration to the region of high transverse momentum, 
where we can perform expansion of the scattering cross section with respect
to the dipole size $r$, the variable conjugate to the transverse momentum.  
We perform  explicit derivation only for quarks scattering off the target, or for the 
fundamental representation of the  Wilson line, Eq.~\eqref{eq:V_rho}.
The final result for the cumulants of an even order~\footnote{For the odd orders one has to be a bit more careful, 
since the adjoint representation of SU(3) is real and thus cannot give rise to odd cumulants. 
For the fundamental representation this is, in principle, possible; however, owing to 
the approximate equality between the number of quarks and antiquarks in the projectile, we 
believe that odd cumulants are of a negligible magnitude.}, however, is true for gluons as well. 
This arises from the fact that representation-dependent Casimir factors cancel out in 
normalized observables as cumulants.  

Performing the gradient expansion for the vector potential
\begin{equation}
A^+ (x^-, \bt\pm\frac{\rt}{2}) \approx A^+ (x^-, \bt)  \pm \frac{\rt}{2} \nabt A^+(x^-,\bt)
\label{Eq:gradient}
\end{equation}
we get to the lowest order in $|\rt|$
\be
{\cal S}_1(\rt,\bt) -1 = \frac{(ig)^2}{2N_c}\tr \left( \rt\cdot {\bf E}(\bt) \right)^2 + {\cal O}(r^3)~,
\ee
where the electric field of the target reads
\begin{equation}
E_i(\bt) = - \partial_i \left( \int dx^- A^+(x^-, \bt) \right). 
\label{Eq:E}
\end{equation}


Analogously,  the $m$-quark S-matrix is given by 
\begin{equation}
{\cal S}_m(\rt_1,\bt_1,\ldots,  \rt_m,\bt_m) -1 =  \left(\frac{(ig)^2}{2N_c} \right)^m \prod_{i=1}^m 
 \tr \left( \rt_i\cdot {\bf E}(\bt_i) \right)^2. 
\label{Eq:Sm}
\end{equation}

In order to proceed with the computations we need to specify 
the field-field correlator, which we adopt from the MV model~\cite{MV}
\begin{equation}
\frac{g^2}{N_c} \langle {E}^a_i(\bt_1) {E}^b_j(\bt_2) \rangle = \frac{1}{N_c^2-1} \delta^{ab}
\delta_{ij} Q_s^2 \Delta(\bt_1-\bt_2),  
\label{Eq:MVcorr}
\end{equation}
where a general form of 
the impact parameter dependence of the correlator $\Delta(\bt)$ with the Fourier image $\tilde{\Delta}(\kt)$
 was assumed.
$\Delta(\bt)$ is normalized such that $\Delta(0) = 1$.   
\subsection{Cumulants of azimuthal anisotropy}
The derivation of the cumulants of azimuthal anisotropy reduce to the 
computation of the azimuthal dependence of the fully connected 
expectation value of the S-matrix for $m$ particles.  
The m-th order cumulant is given by 
\begin{equation}
c_2\{m = 2n\} = \langle \exp\left[i 2 (\phi_1+\phi_2+\cdots+\phi_n - \phi_{n+1} - \phi_{n+2} - 
\cdots - \phi_{2n}) \right] \rangle_\phi,  
\label{Eq:c2m}
\end{equation}
where the average with respect to the angular coordinates is defined by 
\begin{equation}
\langle f(\phi_1, \ldots, \phi_m) \rangle_\phi = 
\frac{ \int d\phi_1 \cdots d\phi_m f(\phi_1, \ldots, \phi_m) \langle S_m (\rt_1,  \ldots, \rt_m) -1  \rangle^{\rm conn.}   }
 {  \int d\phi_1 \cdots d\phi_m  \langle S_m (\rt_1, \ldots, \rt_m) -1  \rangle   }. 
\label{Eq:f}
\end{equation}
The azimuthal angles $\phi_m$ in the laboratory frame characterize
each particle (in our case, a dipole with ${\rt}_m=(r_m \cos\phi_m, r_m \sin\phi_m )$). 

To simplify the notation we introduced the S-matrix averaged with respect to the impact parameters, as follows 
\begin{equation}
S_m (\rt_1,  \ldots, \rt_m) = \frac{1}{S_\perp^m} \int d^2b_1 \cdots  d^2b_m  S_m (\rt_1, \bt_1, \ldots, \rt_m, \bt_m),  
\label{Eq:SInt}
\end{equation}
where $S_\perp$ is the transverse area of the projectile. 
In Eq.~\eqref{Eq:SInt} $\langle S_m (\rt_1, \bt_1, \ldots, \rt_m, \bt_m) -1  \rangle^{\rm conn.}$
denotes the fully connected
contribution to the S-matrix. 
We perform the angular average in the $r$-space, because to this order of the S-matrix expansion
it is equivalent to the 
angular average in the momentum space.

\begin{figure}[t]
\begin{center}
\includegraphics[height=5cm]{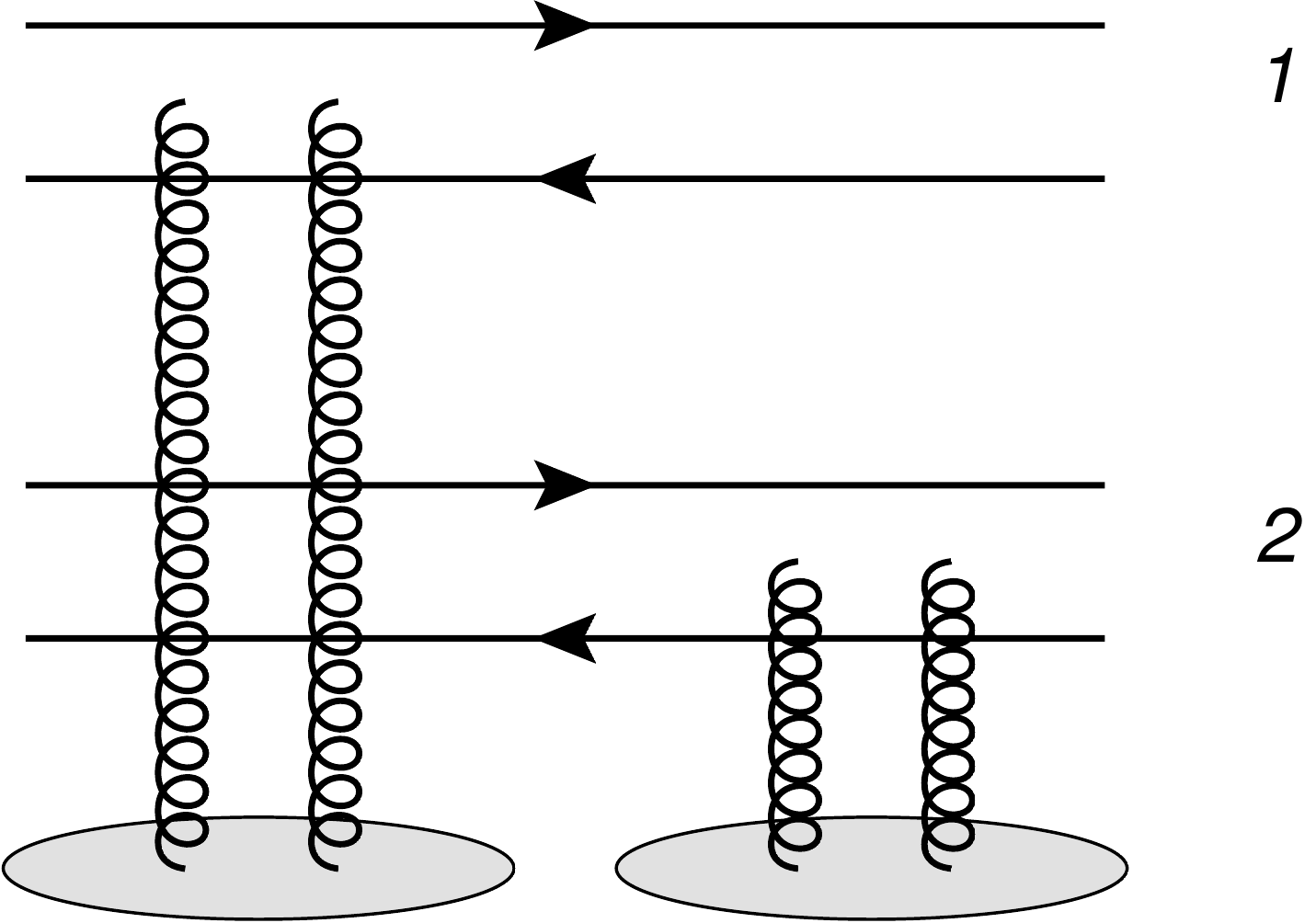}
\hspace{2cm}
\includegraphics[height=5cm]{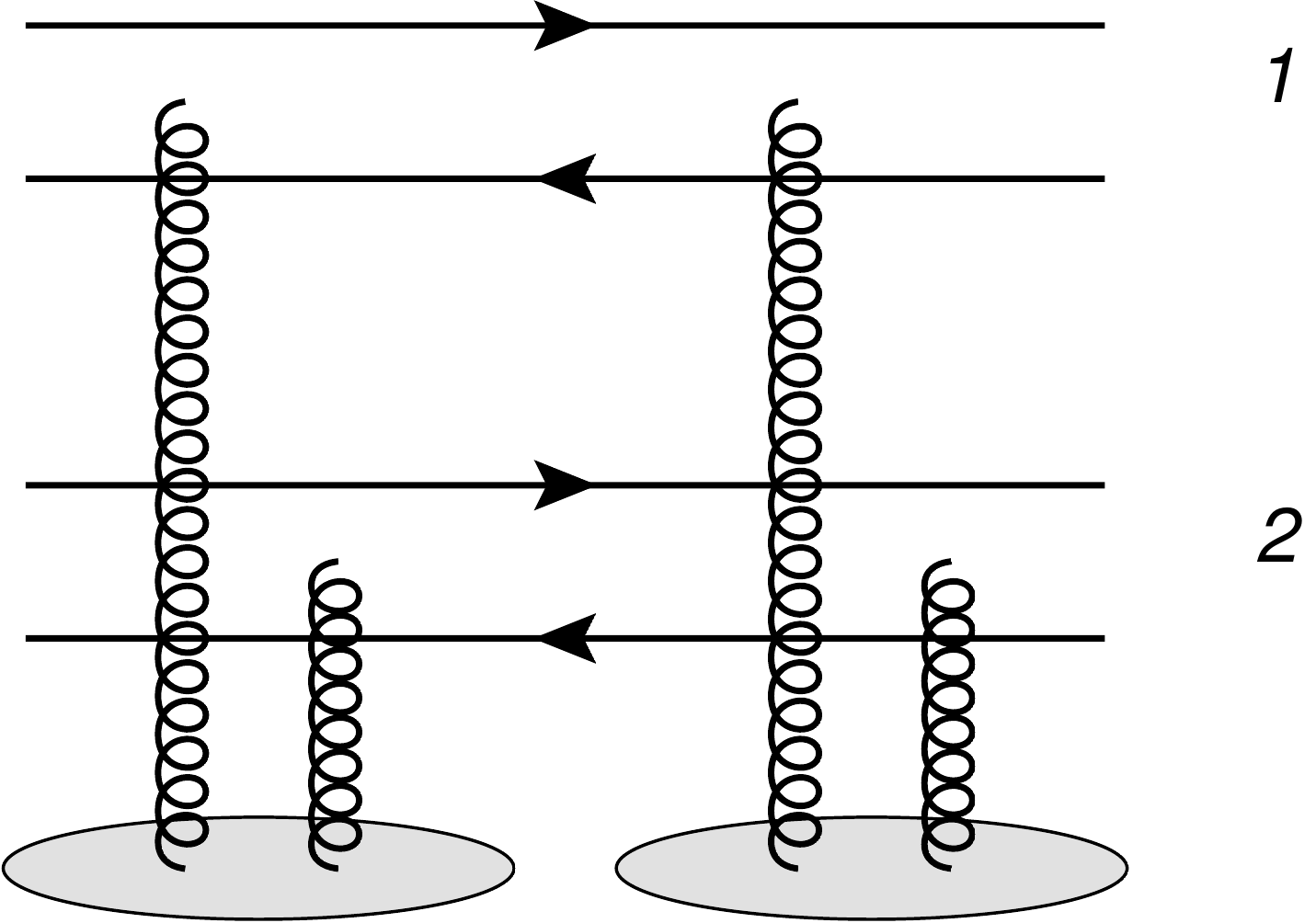}
\end{center}
\caption{ 
The disconnected (left) and connected (right) contributions to the S-matrix for 2 particles.  
} 
	\label{fig:DiscConn}
	\end{figure}

The denominator in Eq.~\eqref{Eq:f} is dominated by the disconnected graph (see Fig.~\ref{fig:DiscConn}) with the corrections suppressed
by powers of $1/N_c$: 
\begin{equation}
\langle S_m (\rt_1, \bt_1, \ldots, \rt_m, \bt_m) -1  \rangle =   \left(\frac{(ig)^2}{2N_c} \right)^m \prod_{i=1}^m 
\langle 
 \tr \left( \rt_i\cdot {\bf E}(\bt_i) \right)^2 \rangle + {\cal O}(N_c^{-2}).   
\label{Eq:denominator}
\end{equation}
Thus to the leading order in $N_c$, the denominator 
\begin{equation}
\langle S_m (\rt_1, \ldots, \rt_m ) -1  \rangle   \approx\left(
-\frac{Q_s^2}{4} \right)^m \prod_{i=1}^m r_i^2.   \label{Eq:denominator1}
\end{equation}

The numerator in Eq.~\eqref{Eq:f} involves all possible contractions that generate the fully connected graphs.  
There are $(2m-2)!!$ ways to contract $S_m (\rt_1, \bt_1, \ldots, \rt_m, \bt_m)$ in a fully connected way. 
Here we show only one term, the rest of  $(2m-2)!!-1$ terms  can be obtained by permutations:
\begin{eqnarray}
\langle S_m (\rt_1, \bt_1, \ldots, \rt_m, \bt_m) -1 \rangle^{\rm conn.} &=& 
\left(\frac{-Q_s^2}{4}\right)^m
\frac{1}{(N_c^2-1)^{m-1}} 
\Delta(\bt_1-\bt_2) \Delta(\bt_2-\bt_1) \cdots \Delta(\bt_{m-1}-\bt_m) \Delta(\bt_m-\bt_1) \nonumber \\ &&
(\rt_1 \rt_2) (\rt_2 \rt_3) \cdots (\rt_{m-1} \rt_m) (\rt_m \rt_1)   + {\rm permutations}. 
\label{Eq:SmConn}
\end{eqnarray}
Averaging with respect to the impact parameter can be best done using Fourier transformation:
\begin{equation}
\frac{1}{S_\perp^m} \int d^2b_1 \cdots  d^2b_m   \Delta(\bt_1-\bt_2) \Delta(\bt_2-\bt_1) \cdots \Delta(\bt_{m-1}-\bt_m) \Delta(\bt_m-\bt_1)  =  \frac{1}{S_\perp^{m-1}} \int \frac{d^2k} {(2\pi)^2} \tilde{\Delta}^m(k).
\label{Eq:AvImp}
\end{equation}
In what follows we consider two correlation functions: a Gaussian
\begin{equation}
\Delta_G(\bt) = \exp\left( - \frac{\bt^2}{\sigma^2} \right)  
\label{Eq:Gauss}
\end{equation}
and an exponential 
\begin{equation}
\Delta_E(\bt) = \exp\left( - \sqrt{2}\frac{b}{\sigma} \right). 
\label{Eq:Exp}
\end{equation}
Both functions are introduced such that 
\begin{equation}
\frac{1}{S_\perp}\int d^2 b \Delta_{G,E}(\bt) = \frac{1}{S_\perp} \pi \sigma^2 = \frac{S^c_{\perp} }{S_\perp} = \xi. 
\label{Eq:sigma}
\end{equation}
Here $\xi$ is the ratio of the correlated area, $S_\perp^c$, to  the area of the projectile, $S_\perp$ (the proton in p-A collisions). 

For these two cases  we have 
\begin{eqnarray}
&&\frac{1}{S_\perp^m} \int d^2b_1 \cdots  d^2b_m   \Delta_G(\bt_1-\bt_2) \Delta_G(\bt_2-\bt_1) \cdots \Delta_G(\bt_{m-1}-\bt_m) \Delta_G(\bt_m-\bt_1)  = \frac{  \xi^{m-1}  }{m}, \\ &&
\frac{1}{S_\perp^m} \int d^2b_1 \cdots  d^2b_m   \Delta_E(\bt_1-\bt_2) \Delta_E(\bt_2-\bt_1) \cdots \Delta_E(\bt_{m-1}-\bt_m) \Delta_E(\bt_m-\bt_1)  = \frac{ \xi^{m-1}  }{3m-2}.
\label{Eq:AvImpFinal}
\end{eqnarray}
Thus after averaging with respect to the impact parameter we obtain 
\begin{eqnarray}
\label{Eq:SmConn1}
\langle S_m (\rt_1, \ldots, \rt_m) -1 \rangle^{\rm conn.} &=&  
\left(\frac{-Q_s^2}{4}\right)^m
\left( \frac{\xi}{N_c^2-1} \right)^{m-1}  
\left\{ 
  \begin{array}{c}
	    m^{-1}  \\
			    (3m-2)^{-1}
					  \end{array} \right\}
(\rt_1 \rt_2) (\rt_2 \rt_3) \cdots (\rt_{m-1} \rt_m) (\rt_m \rt_1) \\ && + {\rm permutations}. \nonumber 
\end{eqnarray}
Here the upper (lower) line corresponds to the Gaussian (exponential) correlator. 
This result can be used to compute the factorial moments along similar lines as for the dense-dense limit, 
see Ref.~\cite{Gelis:2009wh}~\footnote{Indeed a straightforward computation of the factorial moments 
gives $m_q = (q-1)! \frac{K}{2} \left(\frac{\bar{n}}{K}\right)^q \left\{ 
  \begin{array}{c}
	    q^{-1}  \\
			    (3q-2)^{-1}
					  \end{array} \right\}
   $, where $K = (N_c^2-1)/\xi$ and $\bar{n}$ is the average number of scattered partons. Note that this
	 agrees only logarithmically  with the usual negative binomial result: $m^{\rm NBD}_q = (q-1)! K \left(\frac{\bar{n}}{K}\right)^q$, which was obtained in the dense-dense limit in Ref.~\cite{Gelis:2009wh}. }.

\begin{figure}[t]
\begin{center}
\includegraphics[height=5cm]{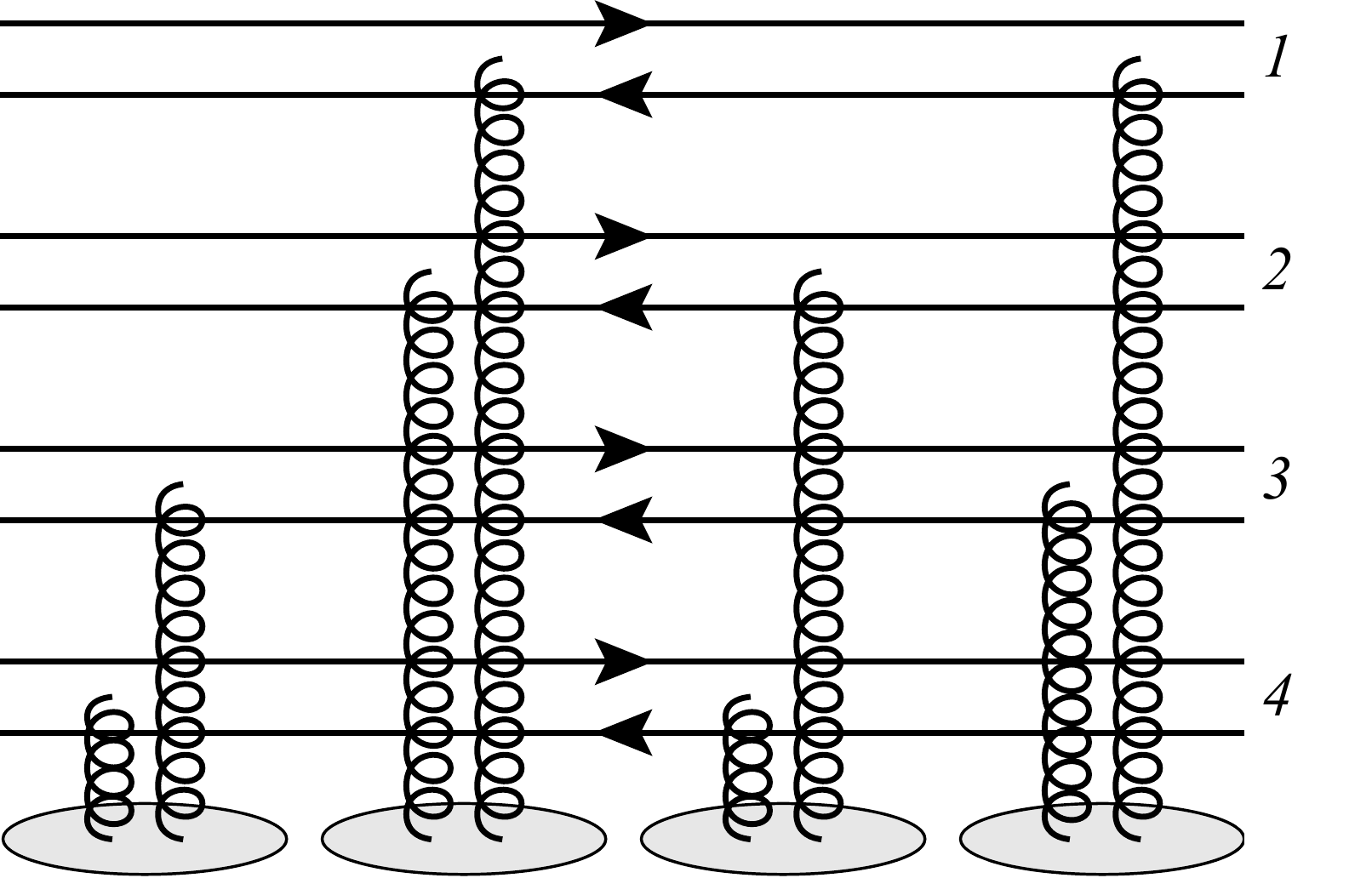}
\hspace{2cm}
\includegraphics[height=5cm]{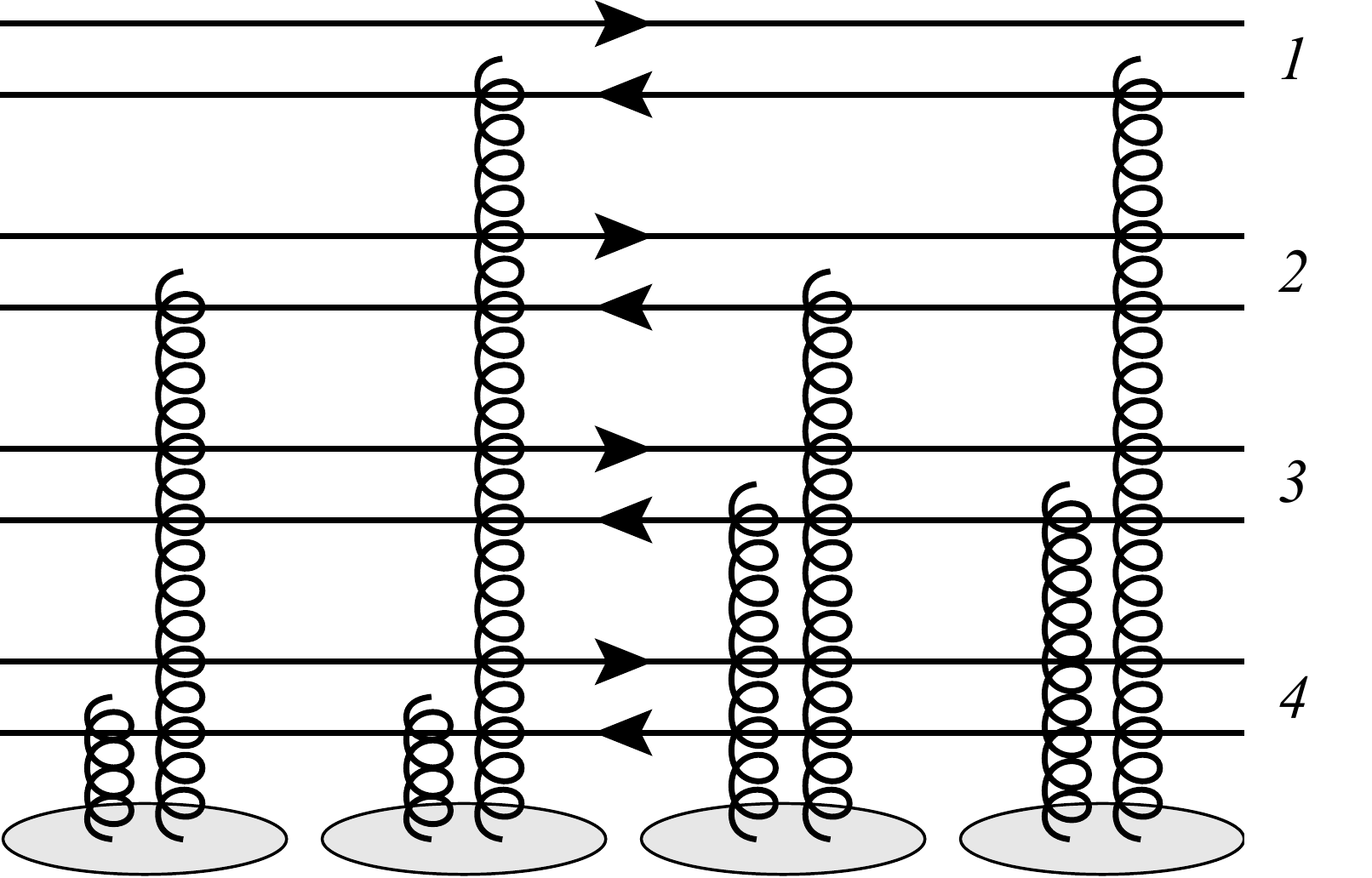}
\end{center}
\caption{ 
An example of two connected contributions to the S-matrix for four particles. 
After performing angular average of $\exp\{2i(\phi_1+\phi_2-\phi_3-\phi_4)\}$ 
only the right diagram provides a non-zero contribution: it involves contractions 
of dipoles 1 or 2  with 3 or 4. 
The left (right) diagram is proportional to 
$(\rt_1\rt_3)(\rt_2\rt_4)(\rt_1\rt_2)(\rt_3\rt_4)$
$\Big((\rt_1\rt_3) (\rt_1\rt_4) (\rt_2 \rt_3) (\rt_2\rt_4)\Big)$. 
} 
	\label{fig:FourP}
\end{figure}

The next step is the integration with respect to all $\phi_i$ of 
$e^{2i(\phi_1+\phi_2+\cdots+\phi_n - \phi_{n+1} - \phi_{n+2} - \cdots - \phi_{2n})}$ weighted 
with $S_m$, where $2n=m$. Not all  the terms in~Eq.~\eqref{Eq:SmConn1} contribute to this average.
In fact, the term we wrote explicitly  down vanishes after the integration.
However, the $m!! (m-2)!!$ nonzero terms remain~\footnote{This is in a quantitative contrast to 
computations of the factorial moments, see Ref.~\cite{Gelis:2009wh} for the dense-dense limit computation of 
the factorial moments.}. 
Those are defined by all possible  contractions of the terms 
entering with opposite signs before $\phi$'s in  $e^{2i(\phi_1+\phi_2+\cdots+\phi_n - \phi_{n+1} - \phi_{n+2} - \cdots - \phi_{2n})}$. An example of a term resulting
in  a nonzero contribution is 
$\propto (\rt_1 \rt_{n+1}) (\rt_1 \rt_{n+2}) (\rt_2 \rt_{n+2}) (\rt_2 \rt_{n+3})  \cdots (\rt_{n-1} \rt_{2n-1})  
(\rt_{n-1} \rt_{2n}) (\rt_{n} \rt_{2n}) (\rt_{n} \rt_{n+1})$. See also an explicit example in Fig.~\ref{fig:FourP}. 
The angular integration of every  such term gives an extra factor of $1/2$. Hence we 
obtain~\footnote{This result can be further simplified by taking into account that for even $m$: $m!! = 2^{m/2} (m/2)! $.}  
\begin{eqnarray}
c^G_2\{m\} &=& \frac{m!! (m-2)!!}{m \ 2^m} \left( \frac{\xi}{N_c^2-1}\right)^{m-1}, \\ 
c^E_2\{m\} &=& \frac{m!! (m-2)!!}{(3m-2) 2^m} \left( \frac{\xi}{N_c^2-1}\right)^{m-1}  
\label{Eq:cFinal}
\end{eqnarray}
for the Gaussian and exponential correlators respectively.  
The harmonics of the azimuthal anisotropy  are related to the cumulants by 
\begin{equation}
v^m_2\{m\} = \kappa_m c_2\{m\},
\label{Eq:vnm}
\end{equation}
where, as shown in Appendix, 
\begin{equation}
\kappa_{2 n} 
=  (-1)^{n+1} \left[{n!(n-1)! \sum_{k=1}^\infty \left( \frac{2}{j_{0,k}} \right)^{2n}}\right]^{-1}. 
\label{Eq:KappaFin_text}
\end{equation}
Here $j_{0,k}$ is the $k$-th zero of Bessel function $J_0(x)$.  
For a large orders $n$ we have 
\begin{equation}
 \kappa_{2 n}  \approx  \frac{(-1)^{n+1}}{n!(n-1)!} \left( \frac{j_{0,1}}{2} \right)^{2n}. 
\label{Eq:betaLarge}
\end{equation}

Finally we obtain for even $m$ 
\begin{equation}
(v_2^G\{m\})^m  = \frac{(-1)^{\frac{m}{2}+1} }{2 m \sum_{k=1}^\infty \left( \frac{2}{j_{0,k}}\right)^{m}} \left( \frac{\xi}{N_c^2-1} \right)^{m-1} \quad {\rm and} \quad  
(v_2^E\{m\})^m  = \frac{(-1)^{\frac{m}{2}+1} }{2(3m-2)  \sum_{k=1}^\infty \left( \frac{2}{j_{0,k}}\right)^{m}   } \left( \frac{\xi}{N_c^2-1} \right)^{m-1}. 
\label{Eq:Final}
\end{equation}
This result also holds for gluons scattering off the target. 

\begin{figure}[t]
\begin{center}
\includegraphics[height=7cm]{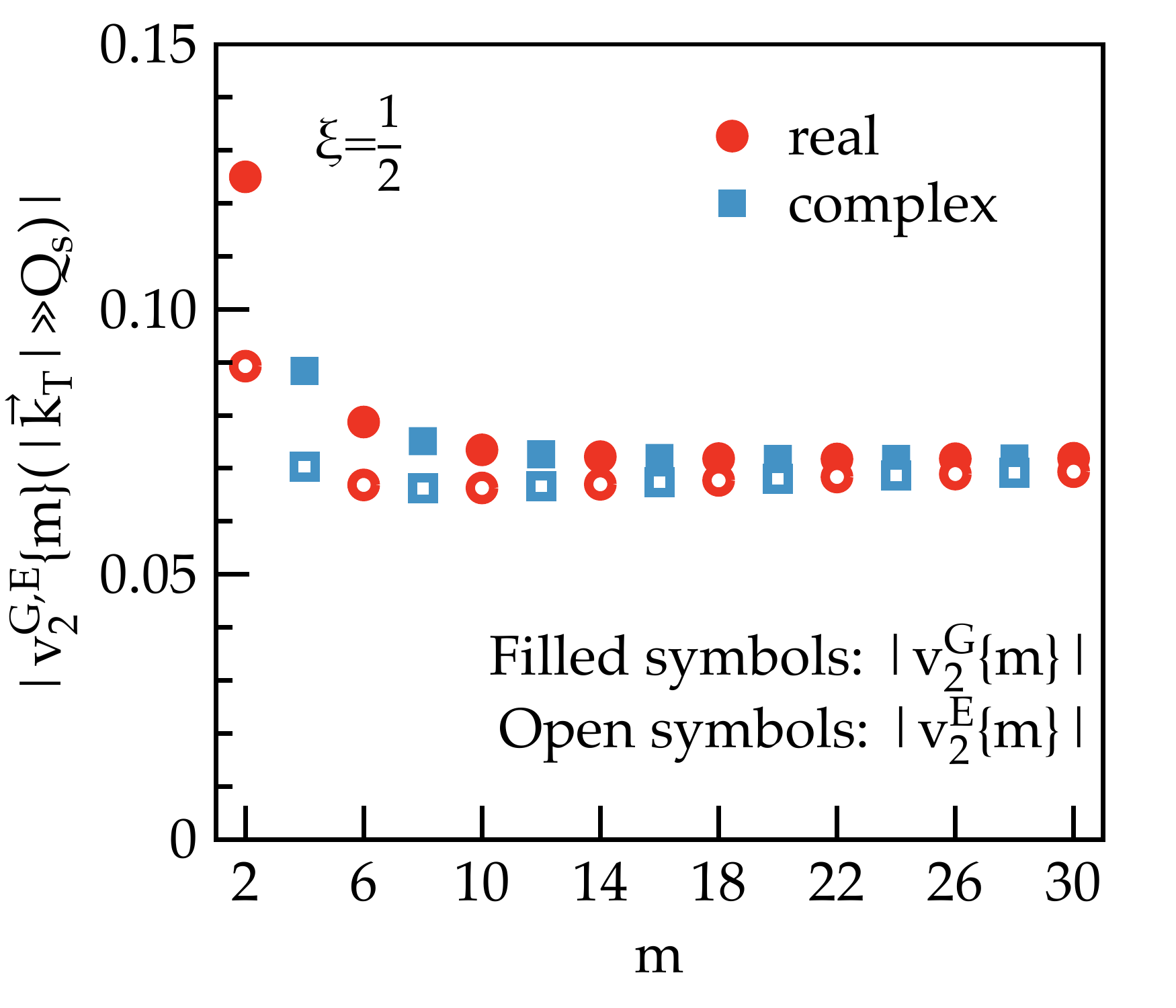}
\end{center}
\caption{ The absolute value of $v_2^{G,E}\{m\}$ as a function of $m$. The filled (open) symbols correspond to 
$|v^G_2\{m\}|$ ($|v^E_2\{m\}|$). The filled and open circles (squares) denote real (complex) $v_2\{m\}$. 
The calculations are 
performed for $\xi=1/2$. Phenomenologically relevant values for $\xi$ can be found elsewhere~\cite{AniCGC}. } 
	\label{fig:1}
	\end{figure}

For a large order $m$ we get for the absolute value of the harmonics 
\begin{equation}
\lim_{m\to\infty} |v_2\{m\}| = \frac{\xi}{N_c^2-1} \frac{j_{0,1}}{2}
\label{Eq:Absolute}
\end{equation}
for both $c^G_2\{m\}$ and  $c^E_2\{m\}$. Although we were unable to prove this rigorously, 
we  believe that this result  holds for any short range $\Delta(\bt)$.  

We established the equality of the absolute values of  high order harmonics. The   
main point of this article is that the fully connected graphs give {\it positive} cumulants of 
{\it any} order and since $\kappa_{m}$
alternate sign depending on the order $m$, every second $v_2\{m\}$ is {\it complex}!  
The lowest order at which $v_2\{m\}$ becomes complex is the {\it fourth} 
\begin{equation}
(v_2^{G,E}\{4\})^4 = -  c_2\{4\} = - \left\{ 
  \begin{array}{c}
	    1/4  \\
			    1/10
					  \end{array} \right\} 
						\left(\frac{\xi}{N_c^2-1} \right)^3 < 0,
\label{Eq:v4}
\end{equation}
as is also illustrated in Fig.~\ref{fig:1}.
One can extend this consideration for the high momentum region of the dense-dense limit,
the so-called ``glasma'' graph, and show that in this case $v_2\{4\}$ is complex as well.  

\section{Discussion} 
As we established in the previous section, while the absolute values of 
high order harmonics are approximately equal to each other and therefore follow the hierarchy observed in 
high-energy p-A collisions ($v_2\{4\} \approx v_2\{6\}  \approx v_2\{8\}   $), every second 
coefficient starting from $v_2\{4\}$ is complex in contradiction to what is seen in experiments at high multiplicities. 
For small multiplicities, however, experimental $c_2\{4\}$ is positive, i.e. $v_2\{4\}$ is complex. 
This implies that the connected graphs considered here may potentially describe only 
low multiplicity collisions. High multiplicities are dominated by other mechanisms. This might include final state interactions
(i.e. rescattering, not related to the initial state CGC dynamics) or an initial state 
effect not accounted for by the formalism considered above.
The former possibility is discussed in detail in  the literature, see 
Refs.~\cite{hydro,Bzdak:2014dia}. The possibility for the latter was put forward by Kovner and Lublinsky in Ref.~\cite{KL} and recently developed 
in Refs.~\cite{AniCGC,AniC4,AniMV,Noronha:2014vva}: the main idea is that the target's electric field being a vector 
necessarily should points to some 
direction of the transverse impact parameter space and form a domain structure of the target. Partons scattering off
this electric field receive the same momentum kick and this generates multi-particle long-range rapidity correlation. 
We stress that  this correlation arises from a single particle  azimuthal anisotropy and thus the higher 
order harmonics
are real if the contribution, defined by the disconnected graphs, dominates over the connected graphs 
(intrinsic  correlations) considered in this paper.
An interested reader is referred to Ref.~\cite{AniC4} for the detailed analysis of the fourth order cumulant with 
connected and disconnected graphs taken into account. 
Recent MV model calculations and JIMWLK  evolution~\cite{AniMV} indeed showed that the mentioned one-particle 
azimuthal anisotropy 
is present in the S-matrix and it does not vanish at small $x$.

Besides the above, the results of this manuscript were derived in the framework of the MV model for the dense target; that is 
we explicitly assumed Gaussian correlations, see Eq.~\eqref{Eq:MVcorr}. As was shown in 
Ref.~\cite{Dumitru:2010mv},  JIMWLK evolution generates higher order correlations, e.g. a four point function 
$\langle E^a E^b E^c  E^d \rangle$  will involve terms of the form
$\frac{1}{N_c} f^{ab e} f^{cd e}$. These corrections  are not necessarily large as shown in Ref.~\cite{Dumitru:2011vk}
 by numerical simulations 
for various higher order function.

\section{Appendix: From cumulants to harmonics of the azimuthal anisotropy}
In this Appendix, we derive a general analytic  relation between 
cumulants and harmonics 
\begin{equation}
\kappa_{2m} = \frac{v^{2m}_n\{2m\}} {c_n\{2m\} } .  
\label{Eq:Kappas}
\end{equation}
We will follow the notation and definitions of  Ref.~\cite{Borghini:2001vi},
including 
the transverse event flow vector $Q$ represented as a complex number, see Eq.~(17) of Ref.~\cite{Borghini:2001vi},
and the definition of the cumulants, Eq.~(12) and  Eq.~(25)  and relation to the ``flow'' harmonics, Eq.~(28) of
Ref.~\cite{Borghini:2001vi}. The latter equation is derived from the  formalism of generating functions and 
defines the connection between the cumulants  
 $\langle\langle |Q|^{2k} \rangle\rangle$ and harmonics $\langle Q\rangle$. For large multiplicity,  
\begin{equation}
\sum \frac{x^{2k}}{(k!)^2} \langle\langle |Q|^{2k} \rangle\rangle = \ln I_0(2x\langle Q\rangle). 
\label{Eq:Generating}
\end{equation}
Expanding the generating equation~\eqref{Eq:Generating} 
up to order $x^{2k}$ and equating the coefficients of $x^{2k}$ one obtains a relation
between the cumulants and harmonics.

Hence the problem  reduces to finding the expansion of 
$ \ln \left( I_0(2 y ) \right)  $ at $y=x \langle Q\rangle =0$. 
The Bessel function 
can be represented as an infinite product
\begin{equation}
I_0(2y) = \prod_{k=1}^\infty \left( 1 + \left(\frac{2y}{j_{0,k}}\right)^{2} \right)
\label{Eq:Iprod}
\end{equation}
and thus 
\begin{eqnarray}
  \ln \left( I_0(2 y ) \right)   &=& 
	\sum_{k=1}^\infty 
\ln\left( 1 + \left(\frac{2y}{j_{0,k}}\right)^2 \right)  = \sum_{i=0}^\infty
 a_i 
y^{2i},
\label{Eq:N2}
\end{eqnarray}
where 
\begin{equation}
a_i =  
\frac{(-1)^{i+1}}{i} 
\sum_{k=1}^\infty 
\left(\frac{2}{j_{0,k}}\right)^{2i}. 
\label{Eq:ai}
\end{equation}
Using the last equation 
we obtain  the wanted relation
\begin{equation}
\kappa_{2m} = \frac{ {\frac{1}{(m!)^2}} }{a_m} 
=  (-1)^{m+1} \left[{m!(m-1)! \sum_{k=1}^\infty \left( \frac{2}{j_{0,k}} \right)^{2m}}\right]^{-1}. 
\label{Eq:KappaFin}
\end{equation}
The sum  $\sum_{k=1}^\infty \left( \frac{2}{j_{0,k}} \right)^{2m}$ can be well approximated by the first term for $m>2$ (see the Table I). 
At large $m$ this becomes a robust limit: 
\begin{equation}
\lim_{m\to \infty} \kappa_{2m} = (-1)^{m+1} \frac{(j_{0,k}/2)^{2m}} { m! (m-1)! }.
\label{Eq:Lim}
\end{equation}
In Table I,  we list first few coefficients $\kappa_{2m}$ and their approximation by Eq.~\eqref{Eq:Lim}.

Since the zeros of the Bessel function, $j_{0,k}$, are not always convenient, we also derived an 
alternative expression by  noticing that the function $f(x)  = \ln \left( I_0(2x) \right)  $
satisfies the following differential equation:
\begin{equation}
f'' +(f')^2 + \frac{f'}{x} -4 =0; \quad {f(0)=0}. 
\label{Eq:f}
\end{equation}
Expanding $f=\sum_n a_n x^{2n} $ and equating the coefficient of the same power of $x$, one can show that 
\begin{equation}
a_i = \frac{(-1)^{i+1}}{2i} \beta_{2i}, 
\label{Eq:a_through_beta}
\end{equation}
where 
$\beta_n$ satisfies the following recursive relation
\begin{equation}
\beta_n = \frac{1}{n} \sum_{i=2}^{n-2} \beta_i \beta_{n-i}; \quad \beta_{2i+1}=0; \quad \beta_2=2.
\label{Eq:beta}
\end{equation}
The coefficients relating the cumulants and harmonics in terms of $\beta_{2n}$ are then  
\begin{equation}
\kappa_{2n} = (-1)^{n+1} \frac{2 } {n! (n-1)! \beta_{2n}}.
\label{Eq:Kappa}
\end{equation}

\begin{table}
\begin{tabular}{|c|c|c|c|c|c|}
\hline Order, $2m$  & 6 & 8& 10 & 12 & 14  \\  
\hline $1/\kappa_{2m}$ & $-4$ &  33 & $-456$ & 9460 & $- 274800$ \\ 
\hline $1/\kappa_{2m}$ from Eq.~\eqref{Eq:Lim} & $- 3.97$& 32.96 & $- 455.886$ & 9459.56 & $- 274797.56$\\ 
\hline 
\end{tabular}
\caption{First few non-trivial coefficient $\kappa_{2m}$ and their approximation. $\kappa_2=\kappa_4=1$.}
\end{table}

\begin{acknowledgments}
I am  grateful to my collaborators,  Adrian Dumitru and Larry McLerran, 
whose input to the manuscript was invaluable. 
I am especially thankful to Adrian Dumitru for encouraging me to write this article. 
I thank Mateusz Ploskon and the organizers of the conference Initial Stages in High-Energy Nuclear Collision
2014 for creating an stimulating environment which motivated me to finish this manuscript.
I also thank Robert Pisarski for valuable comments and careful reading of the manuscript. 
\end{acknowledgments}


\end{document}